# Monolithic axial InGaAs quantum dot emitters in GaAs-based nanowires via Sb-mediated facet engineering

*Hyowon W. Jeong*[1,5,*], *Aris Koulas-Simos*[2], *Imad Limame*[2], *Markus Döblinger*[3], *Sang Kyu Kim*[4], *Chirag C. Palekar*[2], *Jonathan J. Finley*[1], *Stephan Reitzenstein*[2], *and Gregor Koblmüller*[1,2,*]

[1]Walter Schottky Institute, TUM School of Natural Sciences, Technical University of Munich, Garching 85748, Germany

[2]Institute for Physics and Astronomy, Technical University Berlin, Berlin 10623, Germany

[3]Department of Chemistry, Ludwig-Maximilians-University of Munich, Munich 81377, Germany

[4]Walter Schottky Institute, TUM School of Computation, Information and Technology, Technical University of Munich, Garching 85748, Germany

[5]Department of Physics, University of California, Berkeley, Berkeley, California 94720, United States

*E-mail: hyowon.jeong@berkeley.edu (H.W.J.); gregor.koblmueller@tu-berlin.de (G.K.)

ABSTRACT

GaAs-based nanowires hosting active quantum heterostructures provide a promising route toward monolithic integration of single-photon sources on silicon, a key requirement for scalable quantum photonics. However, ultrathin axial quantum-emitter formation is often hindered by facet-dependent growth dynamics and rotational twins, which induce lateral overgrowth and compromise interface abruptness. Here, we develop InGaAs-based quantum emitters by tailoring facet evolution via dilute Sb incorporation, which efficiently suppresses twins and promotes confined axial insertion at the growth-front facet. This approach significantly enhances the probability of obtaining abrupt, few-nanometer-thin quantum dots at the nanowire tip. Single-nanowire optical spectroscopy reveals intense, spatially localized emission from the active region with lifetimes as short as (0.51±0.02) ns, and second-order photon-correlation measurements consistently exhibit pronounced antibunching with $g^{(2)}(0)<0.4$, confirming single-photon emission. These results establish a strong correlation between twin density and axial heterostructure formation, identifying defect control as a key factor in realizing monolithically integrated nanowire single-photon sources.

The realization of scalable quantum photonic technologies requires high-purity, deterministic single-photon sources compatible with silicon (Si)-integrated platforms.[1] While site-controlled quantum dots (QDs) have demonstrated excellent scalability, uniformity, and device integration,[2–4] monolithic growth on Si remains challenging due to the large lattice and polarity mismatch, e.g., between GaAs and Si, requiring polar buffer and strain-reducing layers to suppress defect formation and enable high-quality epitaxy.[5] In this context, free-standing III-V semiconductor nanowires (NWs) have emerged as a powerful bottom-up platform for realizing a variety of nanoscale electronic, optoelectronic, and integrated quantum photonic devices.[6–13] Their compatibility with lattice-mismatched substrates such as Si,[14, 15] as well as the efficient optical mode confinement,[16, 17] and the ability to host axial heterostructures in NWs with strain relaxation conditions well beyond the critical thickness limit of planar structures[18] collectively underpin their versatility.

In particular, quantum heterostructures embedded along the NW axis offer unique opportunities for site-controlled and scalable single-photon sources, combining strong carrier confinement with efficient photon extraction along the NW waveguide.[17, 19, 20] Furthermore, NW-based QDs grown along the [111] direction can exhibit highly symmetric confinement potential, which suppress the fine-structure splitting (FSS) arising from structural or compositional asymmetries.[21] The near-elimination of FSS makes such NW-QDs promising platforms for the generation of polarization-entangled photon pairs.[22, 23]

Recent progress in III-V heterostructure NW-QD platforms has revealed highly promising quantum emission performance, particularly in devices realized through droplet-assisted vapor-liquid-solid (VLS) growth. InP/InAsP NW-QDs, for instance, have achieved nearly ideal single-photon purity, with $g^{(2)}(0) < 0.05$,[24–26] improved photon indistinguishability,[25, 27] as well as high brightness[25] and high degree of entanglement.[23] AlGaAs/GaAs NW-QDs have likewise shown decent single-photon purity with $g^{(2)}(0) = 0.19$ together with high brightness levels.[28] However, conventional VLS growth, which relies on a liquid droplet (e.g., Au or self-induced group-III elements), often encounters droplet-related limitations in morphology and structural uniformity, and abrupt heterointerface formation,[29–32] while foreign metal catalysts can additionally introduce contamination into the device structure.[33]

Catalyst-free vapor-solid (VS) growth via selective-area epitaxy (SAE) offers several key advantages for quantum emitters in monolithic NW devices, given the droplet-free growth and compatibility with complementary metal-oxide-semiconductor (CMOS) technology for integration on Si platforms. This approach has demonstrated well-defined axial quantum heterostructures in various III-V NW material systems, including GaAs/InGaAs, GaAs/AlGaAs, and GaAs/GaAsP.[34–45] Here, GaAs/InGaAs-based axial NW-QD emitters are particularly attractive due to their excellent optical properties,[34–39] while the InGaAs materials system also offers potential for single-photon sources at telecommunications wavelengths.[1, 46] Achieving the precise thickness control and abrupt interfaces required for well-defined GaAs/InGaAs QDs or quantum disks remains still a key challenge toward deterministic single-photon sources based on such axial inserts.

Bottom-up GaAs-based NWs naturally grow along the epitaxial [111]B direction with sidewall facets and other minor inclined facets near the growth front that belong to the {-1-10} family of planes. A major obstacle in tailoring the size and shape of axially embedded QD emitters arises from the complex, facet-dependent morphology evolution inherited from the underlying (111)B GaAs-based NW core.[47–51] In the so-called twin-induced growth mechanism, which also governs InGaAs segment formation,[45] a strong interplay between facet-dependent growth kinetics and rotational twin formation exists that critically shapes the evolution of the NW growth front.[49, 50] In particular, at high twin densities, subsequent layer deposition tends to enhance lateral infilling over the inclined {-1-10} facets adjacent to the NW growth front.[47, 50] During InGaAs-based segment insertion, this growth mode results in unintended preferential deposition on these inclined facets rather than on the flat (111)B top facet, thereby reducing growth selectivity.[44] Consequently, structural and compositional nonuniformities arise in the segment, hindering the realization of ultrathin axial quantum structures with abrupt heterointerfaces and well-defined emission characteristics.

Previous studies have shown that a small molar fraction (≈2–3%) of antimony (Sb) acts as an effective surfactant during catalyst-free GaAs-based NW growth, modifying surface reconstruction, crystal-phase stability, and adatom diffusion dynamics, which in turn significantly improve the structural and optical properties.[52–54] Notably, precise tuning of the Sb content was found to be decisive in tailoring the rotational twin density,[53] which is essential for regulating facet evolution at the growth front and restoring growth selectivity.

Here, we exploit Sb surfactant-induced twin suppression in catalyst-free SAE growth to realize ultrathin InGaAs-based quantum-confined heterostructures and QDs embedded in GaAs-based NWs (hereafter referred to as InGaAs(Sb) and GaAs(Sb), respectively). By confining the axial insertion thickness to the few-*nm* length scale through Sb-mediated facet engineering, we establish a defect-minimized growth window that yields InGaAs(Sb) segments with abrupt compositional and structural transitions at the heterointerface. Structural analysis by scanning transmission electron microscopy (STEM) combined with energy-dispersive X-ray spectroscopy (EDXS) reveals a strong correlation between axial insertion characteristics and twin defect density. Cathodoluminescence (CL) and micro-photoluminescence (μPL) measurements, along with Hanbury Brown and Twiss (HBT) experiments on individual NWs are further used to investigate the emission in terms of intensity and spatial localization as well as the single photon emission character from the InGaAs(Sb) active region at the NW tip. Second-order photon-correlation $g^{(2)}(\tau)$ measurements exhibit pronounced antibunching, confirming the QD nature of the InGaAs(Sb) region and its suitability as a single-photon source.

**Figure 1**a schematically displays the designed NW heterostructure, in which an InGaAs(Sb) QD emitter is axially embedded within a GaAs-based NW grown directly on a SAE-prepatterned $SiO_2$/Si(111) substrate using molecular beam epitaxy (MBE). As detailed in **S1. Methods** (Supporting Information), a small fraction of Sb (≈3–4 %) is incorporated into both the GaAs(Sb) core and the InGaAs(Sb) insert, while the shell consists of $Al_{0.3}Ga_{0.7}As$/GaAs surface passivation layers.

The magnified schematic in **Figure 1**b illustrates the underlying growth concept. At the growth front of the GaAs(Sb) core, three inclined {-1-10} facets, associated with crystal symmetry of the underlying zinc-blende (ZB) structure, emerge from the corners and converge toward the center, forming a triangular plateau on the (111)B top facet.[49, 50] The presence of these inclined facets, together with facet-dependent growth kinetics, governs subsequent material deposition.[47–51] The addition of dilute Sb during InGaAs growth is intended to suppress rotational twin formation, thereby enhancing the probability of axial growth on the (111)B top facet of the underlying NW core (i), while minimizing unintended deposition on the {-1-10} inclined facets (ii), which is crucial for achieving a well-defined QD emitter.

**Figure 1**c presents a representative scanning electron microscopy (SEM) image of the as-grown NW array. Across the investigated field, the NWs exhibited a high yield (≈90%) and

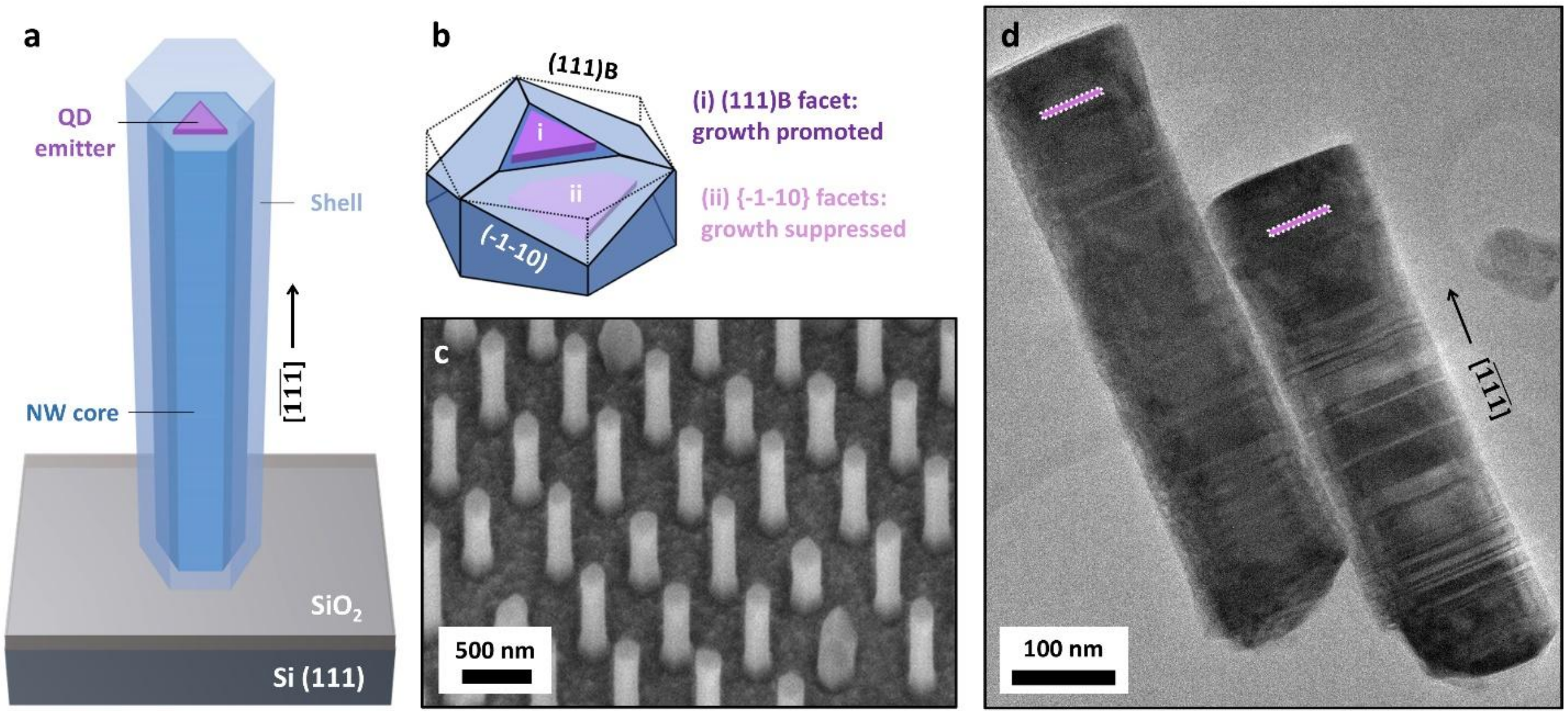

**Figure 1.** Growth concept and structural overview of axially embedded QDs in NWs. (a) Schematic illustration of an InGaAs-based QD axially embedded within a GaAs-based NW directly grown on a SAE-prepatterned $SiO_2$/Si(111) substrate. (b) Magnified schematic of the InGaAs(Sb) region, where axial growth is promoted on the (111)B top facet of the underlying NW core (i) while undesired growth on the {-1-10} inclined facets is suppressed (ii). (c) Representative SEM image of the as-grown NW arrays. The image was taken at 45° and tilt-corrected to display actual lengths. (d) TEM micrograph of representative NWs with the intended InGaAs(Sb) regions highlighted.

uniform morphology, with consistent lengths of $L$ = (580±40) nm and diameters of $D$ = (150±10) nm. **Figure 1**d shows a TEM micrograph of two NWs mechanically transferred from the same SAE field onto TEM grids, where the intended InGaAs(Sb) insertion region is indicated by purple triangles.

To assess the axial growth characteristics of the InGaAs(Sb) insert, high-angle annular dark-field STEM (HAADF–STEM), along with associated EDXS, was performed on multiple representative NWs (see **S1. Methods**, Supporting Information). **Figure 2**a presents a HAADF-STEM micrograph of the top region of one NW, where the contrast reflects atomic number ($Z$) variations. The axial AlGaAs/GaAs shell deposition is clearly distinguished, with the Al-containing layer appearing darker due to its lower average atomic number. Directly beneath this layer, the InGaAs(Sb) region (yellow arrow) exhibits brighter contrast relative to the surrounding material due to In incorporation. A high-resolution micrograph in **Figure 2**b further resolves the region of interest at the NW tip, encompassing the InGaAs(Sb) insertion. Structural analysis reveals a phase-pure ZB domain free of rotational twins across the entire insert thickness, in line with the anticipated surfactant role of dilute Sb in microstructure control.

The corresponding EDXS elemental map of In, shown in **Figure 2**c, reveals a distinct In enrichment localized at the (111)B top facet, forming a thin, axially confined region with abrupt heterointerfaces. In contrast, radial In incorporation remains negligible, with the measured atomic fraction below the detection limit (< 1%). This preferential InGaAs(Sb) deposition at the top facet is consistent with Sb-mediated suppression of twin formation during growth.

In addition, **Figure 2**d displays the axial compositional line profile acquired through the center of the In-enriched region, as indicated by the white arrow in **Figure 2**c. The In profile (purple line) shows that the maximum In content reaches ≈17.6% within the segment, with a full width at half maximum (FWHM) of ≈4 nm, followed by a rapid decay to background levels. This narrow apparent peak width, which is comparable to the effective spatial resolution of the EDXS measurement (≤ 5nm), supports the formation of a few-*nm*-thin insert with abrupt interfaces. The Sb molar fraction (orange line) reaches ≈3–4% within the same segment, as targeted, and remains at a comparable level extending into the GaAs(Sb) core. These axial growth characteristics of InGaAs(Sb) are in marked contrast to those typically observed for Sb-free InGaAs segments.**[44, 45]** As described earlier, Sb-free InGaAs predominantly deposits on the {-1-10} inclined facets rather than on the (111)B top facet of the core, and these segments exhibit a high twin density (≈1.5 $nm^{-1}$), as shown in **Figure S2** (Supporting Information).

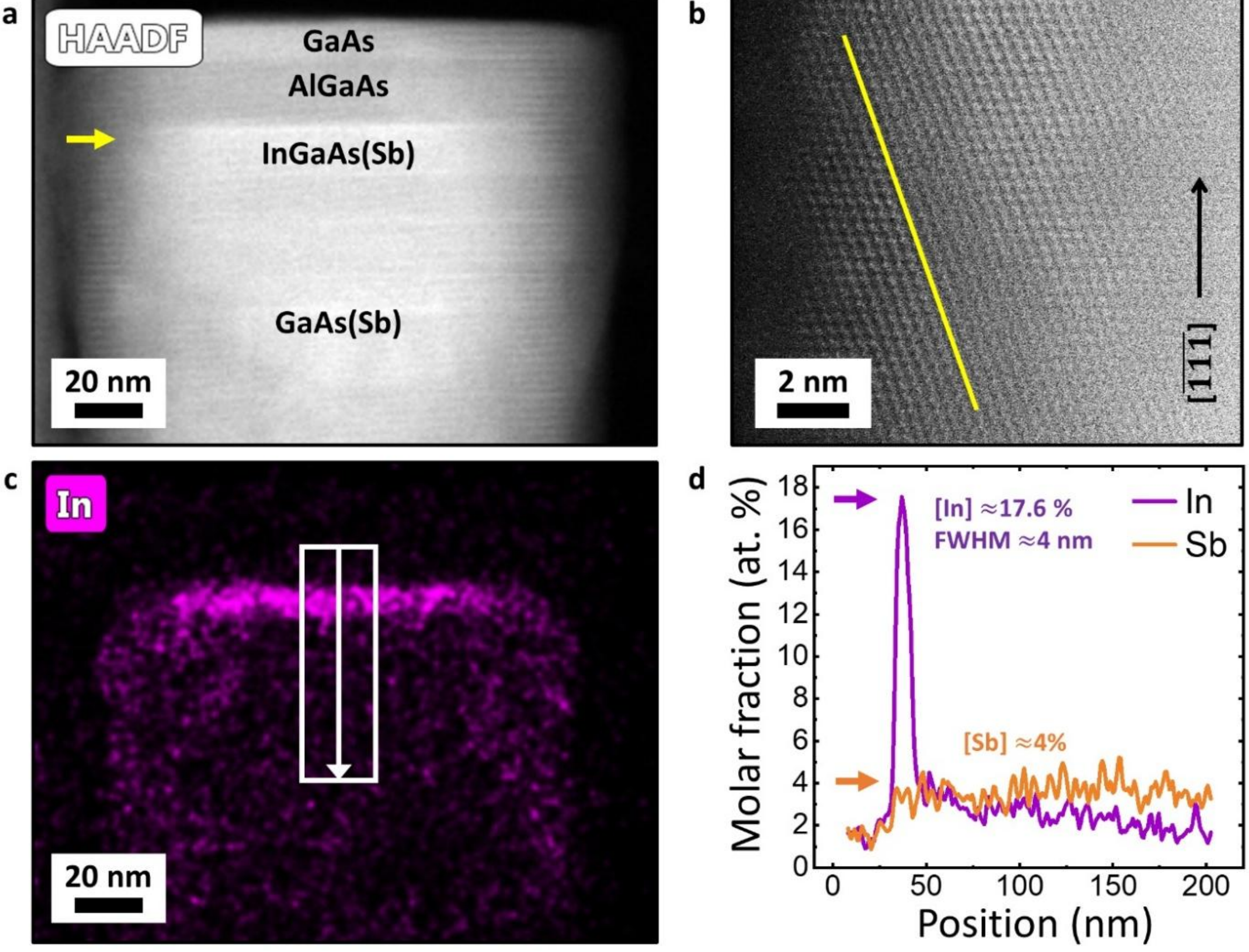

**Figure 2.** Microstructural and compositional analysis of twin-free InGaAs(Sb) axial inserts. (a) HAADF-STEM micrograph recorded in the top region of a transferred NW. (b) High-resolution micrograph magnifying the InGaAs(Sb) region (indicated in (a) by the yellow arrow), with a twin-free ZB domain outlined in yellow. (c) Associated EDXS elemental map of the In distribution, showing an In-rich segment. (d) Vertical compositional profile scanned across the In-rich segment, as indicated by the white arrow in (c). Purple and orange solid lines represent the In and Sb molar fractions, respectively. This plot shows twice the measured atomic fraction of each group III and V element, directly representing the corresponding alloy compositions.

This distinct behavior can be understood by considering two concurrent and largely independent effects governing the axial growth characteristics of the InGaAs-based segment: (i) stabilized growth on the (111)B top facet, which can be attributed to the Sb surfactant effect,[52, 53] and (ii) growth on the inclined {-1-10} facets, which has mainly been associated with the formation of rotational twins.[47–51] In the absence of Sb, frequent twin formation together with a less stable axial growth front favors preferential growth on the inclined facets. However, when dilute Sb is introduced, stabilization of the (111)B growth front combined with the reduced twin probability may act synergistically to promote the formation of axially confined InGaAs(Sb) disks.

Importantly, however, the formation of an axially confined thin disk is not fully deterministic, as twin formation in GaAs-based NWs is inherently stochastic. As shown in **Figure S3** (Supporting Information), a subset of NWs within the same array still exhibits variations in the InGaAs(Sb) deposition profile, with partial incorporation occurring on both the top and inclined facets. In these NWs, a twin density of $\approx 0.5$ $nm^{-1}$ is observed, which is about threefold lower than in Sb-free InGaAs, although twins are not completely eliminated. Under such conditions, growth on the (111)B top facet remains active, but twin-mediated processes simultaneously lead to additional incorporation on the inclined {-1-10} facets, resulting in an overall slightly larger InGaAs(Sb) segment volume. While the limited number of NWs analyzed by TEM precludes a statistically rigorous determination of the fraction of twin-free InGaAs(Sb) inserts across the NW array, the reduced twin density consistently correlates with enhanced axial growth selectivity among the investigated NWs. Further systematic tuning of Sb incorporation and the associated growth conditions may reduce the overall twin density,[53] thereby enabling the system to approach a statistically deterministic regime for axial QD formation.

To verify the spatial origin and emission properties of the InGaAs(Sb) insertion, both CL and μPL measurements were performed at low temperature (20 K and 4 K, respectively) on transferred single NWs (see **S1. Methods** in the Supporting Information). It should be noted that these NWs were transferred onto different substrates ($SiO_2$/Au/Si and sapphire) from those used for TEM structural analysis (copper grids), and therefore, a direct one-to-one structural-optical correlation cannot be established at this stage. Nevertheless, consistent trends observed across multiple NWs provide reliable insight into their general emission characteristics.

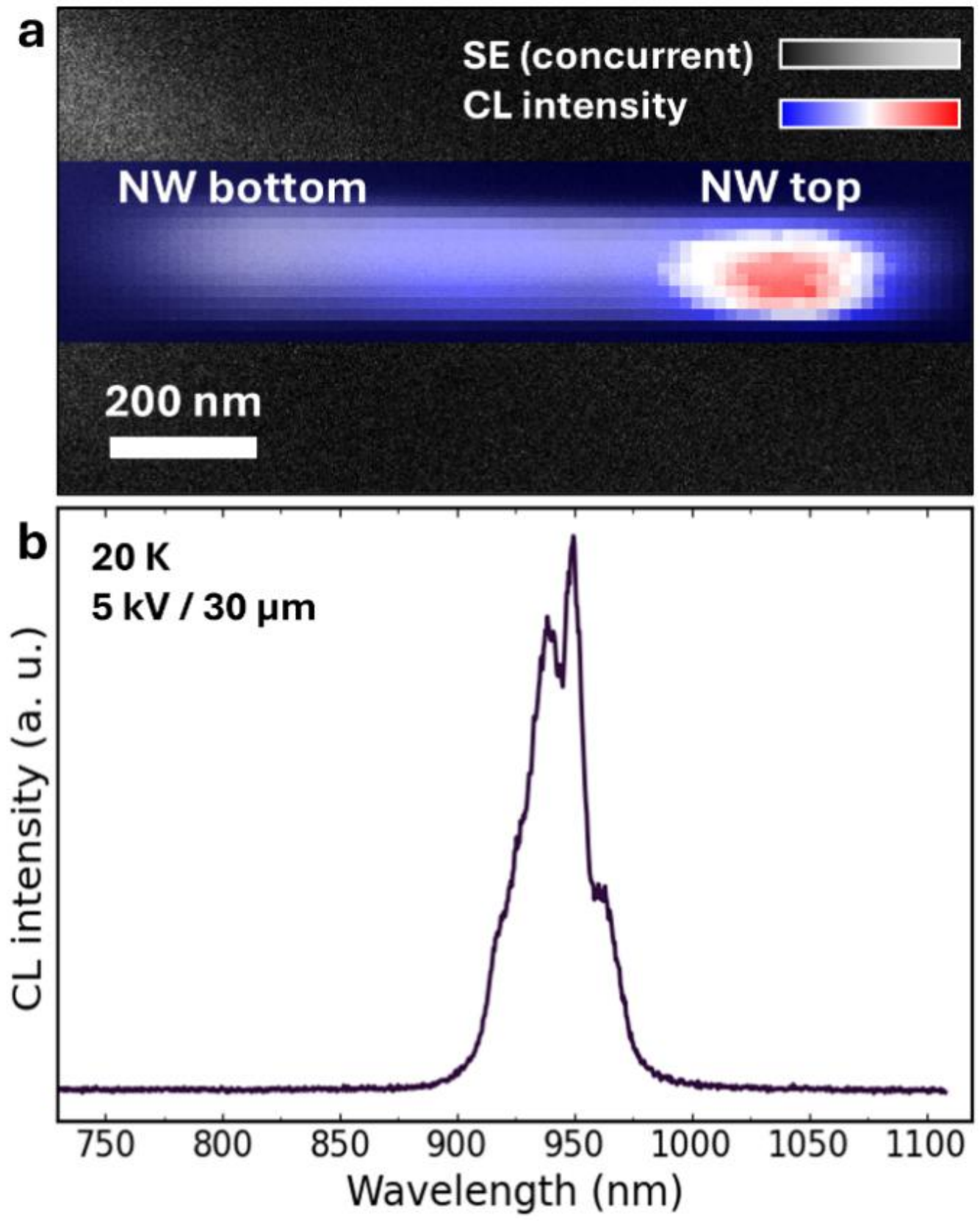

**Figure 3.** CL measurement of a single InGaAs(Sb) NW-QD transferred on a $SiO_2$/Au/Si substrate. (a) CL intensity map overlaid on the concurrently recorded SE image acquired at 20 K using a 5 kV electron beam and 30 µm beam aperture, showing emission localized at the NW top region corresponding to the InGaAs(Sb) active segment, and (b) The corresponding CL spectrum acquired from the same NW, which exhibits an emission band centered at ≈938 nm.

**Figures 3**a,b display a CL intensity map of a representative single NW transferred onto a $SiO_2$/Au/Si substrate, overlaid on the concurrently acquired secondary-electron (SE) image recorded at 20 K using a 5 kV electron beam, along with the corresponding CL spectrum measured from the same NW. The CL emission is strongly localized at the NW top region, confirming that the luminescence predominantly originates from the InGaAs(Sb) active segment.

Note, in CL the density of free carriers is substantially higher than in μPL measurements, leading to enhanced spectral diffusion and less pronounced single-emitter features. Accordingly, the detailed spectral features and possible origin of the observed multi-peak emission are discussed below based on the higher resolution μPL spectra.

**Figures 4**a,b show excitation power-dependent μPL spectra of representative individual NWs under pulsed excitation. The NWs were transferred onto $SiO_2$/Au/Si (a, NW1) and sapphire (b, NW2) substrates, respectively. The $SiO_2$/Au/Si substrate primarily serves as a platform for CL measurements, allowing the μPL characteristics to be correlated with the spatial emission origin from the same NWs, while sapphire was also employed as a favorable substrate for high-quality μPL measurements. For NWs on both substrates, emission from the InGaAs(Sb) insertion was consistently observed. In the representative spectra shown here, the highest-intensity peak is located near ≈1002 nm for the NW1 and ≈952 nm for the NW2 sample, with a linewidth (FWHM) of ≈2 nm in both cases. These spectrally well-defined peaks were selected for the corresponding time-resolved PL (TRPL) and second-order photon-correlation measurements discussed below. Although the dominant emission wavelengths differ between the two NWs, both remain within the range expected from the estimated quantum-confined bandgap energy (≈1.25–1.39 eV, corresponding to ≈890–990 nm), obtained from calculations based on the measured dimensions and composition of the InGaAs(Sb) insertion (see **Section S4** of the Supporting Information). Additionally, the characteristic GaAs(Sb) core emission, expected around 850 nm at low temperature,[44, 53] was completely absent across the investigated NWs, supporting highly efficient transfer to the InGaAs(Sb) region despite its small volume fraction within the NW (< 0.5%).

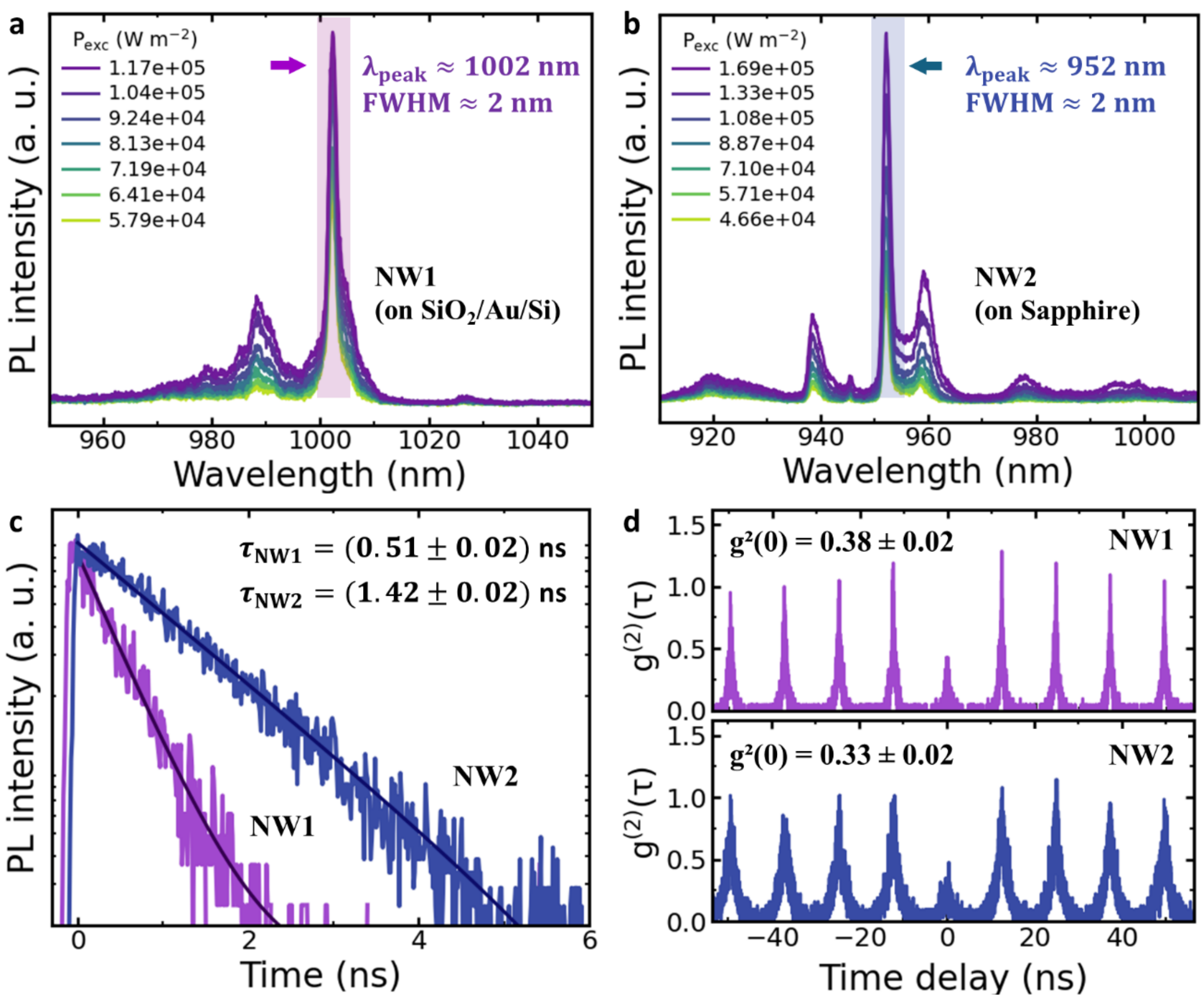

**Figure 4.** μPL analysis of single InGaAs(Sb) NW-QDs. (a,b) Excitation power-dependent μPL spectra of single NWs transferred onto $SiO_2$/Au/Si (a, NW1) and sapphire (b, NW2) substrates at 4 K, exhibiting emission from the InGaAs(Sb) insertion. (c) TRPL decay curves measured from NW1 (purple) and NW2 (dark-blue), along with single exponential fits yielding lifetimes of (0.51±0.02) ns and (1.42±0.02) ns, respectively. (d) Pulsed second-order photon-correlation measurements $g^{(2)}(\tau)$ of the highest-intensity emission peak from NW1 (top) and NW2 (bottom). The pronounced suppression of the zero-delay peak confirms the antibunched photon statistics of single photon emission, with $g^{(2)}(0)$ of 0.38±0.02 and 0.33±0.02, respectively.

Notably, each InGaAs(Sb) spectrum shows a multi-peak structure with relatively broad emission features, rather than a single narrow emission line exhibiting FWHM values below several tens of μeV, as observed in other optimized NW-QD systems.[24–26, 28] This behavior may arise from compositional fluctuations of In and Sb within the InGaAs(Sb) segment, morphological variations during axial deposition (e.g., growth on both the top and inclined facets), and differences in the local crystal structure (e.g., twin density) as noted above. For example, **Section S5** of the Supporting Information provides correlated CL data acquired from the NW1 sample, suggesting that different spectral features can originate from spatially distinct regions within the same NW. More broadly, the majority of the investigated NWs tended to exhibit dominant emission in one of two spectral regimes, centered near ≈1000 nm or ≈950 nm. This distribution may be attributed to a coupled interplay between twin-density-dependent growth facet formation, facet-dependent Sb incorporation, and the influence of Sb incorporation on twin formation. The detailed relationship between these effects and their impact on the optical properties provides scope for further investigation.

Additionally, **Figure 4**c shows TRPL decay curves obtained from NW1 (purple) and NW2 (dark-blue), which reveal lifetimes of (0.51±0.02) ns and (1.42±0.02) ns, respectively. In particular, the fast-decaying emitter with ≈1000 nm wavelength emission is representative for lifetimes typical of standard Stranski-Krastanov (SK)-type InGaAs QDs.[5]. The pronounced variation in lifetime between different NW-QDs is consistent with the overall emitter-to-emitter variability observed in the μPL spectra, and may reflect differences in local structure and composition, such as In content, segment size or shape, and twin defect density.

To assess the quantum nature of the emission, **Figures 4**d shows the second-order photon-correlation functions $g^{(2)}(\tau)$ measured from the selected peak of NW1 (top) and NW2 (bottom), respectively. The measurements were performed under pulsed excitation using a HBT autocorrelation setup integrated with the μPL system (see **S1. Methods** in the Supporting Information). Photon antibunching behavior is reflected by the suppression of the center peak at zero-time delay in $g^{(2)}(\tau)$ compared to uncorrelated peaks at $\tau \gg 0$, indicating sub-Poissonian photon statistics characteristic of nonclassical emission.

The pronounced suppression of the central peak, yielding $g^{(2)}(0)$ of 0.38±0.02 for NW1 and 0.33±0.02 for NW2, unambiguously demonstrate the single-photon emission (with well-suppressed multi-photon error) from the individual InGaAs(Sb) QDs under the small mean photon number assumption. While an ideal single-photon source would exhibit $g^{(2)}(0)$ approaching zero, as demonstrated by the near-ideal single-photon purity with $g^{(2)}(0) < 0.05$ reported for optimized NW-QD systems,[24–26] the relatively large values obtained here can be attributed not only to spectral overlap from adjacent emission lines and residual background emission, but also to the unintended structures discussed above and the associated variations in their shape, size, and alloy composition, together with the relatively large lateral dimensions of the NWs. Nevertheless, the $g^{(2)}(0)$ values measured from multiple NW samples across different substrates remain well below the conventional threshold of 0.5 for single-photon emitters; for example, even lower values were obtained for selected emitters, including a best raw $g^{(2)}(0)$ of 0.27±0.08 and a temporally post-selected value of 0.12±0.07, as shown in **Section S6** of the Supporting Information. While important challenges remain, including elucidating and controlling the origin of the emission wavelength variability, improving the spectral characteristics, and achieving more deterministic formation of the QD active region, these results nevertheless highlight the strong potential of axial InGaAs(Sb) NW-QDs as monolithic single photon sources.

In summary, we have realized ultrathin axial InGaAs(Sb) QD-based single photon sources embedded in GaAs-based NWs via catalyst-free SAE growth. By exploiting Sb-mediated suppression of rotational twin formation, axial QD growth on the (111)B top facet is promoted while unintended deposition on the {-1-10} inclined facets is significantly reduced. Structural analysis demonstrates the formation of a few-*nm*-thin InGaAs(Sb) region with abrupt compositional transitions and a twin-free ZB domain across the insertion thickness. Single-NW optical spectroscopy reveals intense, spatially localized emission from the InGaAs(Sb) QD and fast lifetimes of as short as (0.51±0.02) ns. Second-order photon-correlation function $g^{(2)}(\tau)$ measurements under pulsed excitation consistently exhibit pronounced antibunching, with raw $g^{(2)}(0)$ values below 0.4 and a lowest post-selected value of 0.12±0.07, confirming the single-photon nature of the emission despite their still large lateral dimensions and therefore relatively weak lateral confinement. Although the formation of fully twin-free axial inserts remains stochastic, the observed correlation between reduced twin density and preferential axial growth highlights the critical role of surfactant-mediated facet control. Our findings establish Sb-assisted SAE growth as a viable route toward monolithically integrated axial QDs in GaAs-based NWs directly grown on Si-compatible platforms and provide a foundation for further optimization toward statistically deterministic single-photon sources. Future efforts aimed at NW geometry engineering for enhanced carrier confinement, optimized Sb incorporation and associated growth conditions for more deterministic axial QD formation, and integration with on-chip Si photonic waveguides may enable improved optical QD characteristics, controlled emitter alignment, and Purcell-enhanced optical coupling,[17] advancing scalable NW-based quantum photonic circuits.

## Author Contributions

H.W.J. and G.K. conceived this study. H.W.J. performed MBE growths, SEM measurements and analysis, and sample preparations for experiments. M.D. carried out STEM and EDXS measurements, and H.W.J. analyzed the resulting data. H.W.J. and S.K.K. conducted preliminary μPL and $g^{(2)}(\tau)$ measurements for initial evaluation, and A.K.-S., C.C.P. and I.L. performed comprehensive μPL, TRPL, CL, and $g^{(2)}(\tau)$ measurements. G.K., S.R., and J.J.F. acquired funding, supervised the project, and provided resources. H.W.J. wrote the manuscript with contributions from all authors, and all authors approved the final version of the manuscript.

## Acknowledgement

The authors gratefully acknowledge support from the European Research Council (ERC project QUANtIC, ID: 771747), and the Deutsche Forschungsgemeinschaft (DFG) via Germany's Excellence Strategy-EXC-2111-390814868 (Munich Center for Quantum Science and Technology, MCQST). H.W.J. further acknowledges support by Microelectronics Science Research Center (Nanoscale Hybrids: A New Paradigm for Energy-efficient Optoelectronics) under Contract DE-AC02-05-CH11231. TUB acknowledges funding by the Senate of Berlin, within the Program for the Promotion of Research, Innovation and Technology (ProFIT) cofinanced by the European Regional Development Fund (ERDF, Application No. 0206824, SQALE) as well as by Berlin Quantum (BQ). The authors also thank Hubert Riedl for technical support with MBE, Bianca Seyschab for assistance with preliminary optical measurements, and Hamidreza Esmaielpour for support with sample transfer.

Supporting Information for

# Monolithic axial InGaAs quantum dot emitters in GaAs-based nanowires via Sb-mediated facet engineering

*Hyowon W. Jeong[1,5,*], Aris Koulas-Simos[2], Imad Limame[2], Markus Döblinger[3], Sang Kyu Kim[4], Chirag C. Palekar[2], Jonathan J. Finley[1], Stephan Reitzenstein[2], and Gregor Koblmüller[1,2,*]*

[1]Walter Schottky Institute, TUM School of Natural Sciences, Technical University of Munich, Garching 85748, Germany

[2]Institute for Physics and Astronomy, Technical University Berlin, Berlin 10623, Germany

[3]Department of Chemistry, Ludwig-Maximilians-University of Munich, Munich 81377, Germany

[4]Walter Schottky Institute, TUM School of Computation, Information and Technology, Technical University of Munich, Garching 85748, Germany

[5]Department of Physics, University of California, Berkeley, Berkeley, California 94720, United States

*E-mail: hyowon.jeong@berkeley.edu (H.W.J.); gregor.koblmueller@tu-berlin.de (G.K.)

## S1. Methods

*Nanowire (NW) Growth via Selective-Area Molecular Beam Epitaxy (SAE)*

The growth of NWs was performed using a solid-source Gen-II molecular beam epitaxy (MBE) system, equipped with conventional effusion cells for group-III elements (In, Ga, Al) and Veeco valved cracker cells for group-V elements (As, Sb). The As species were supplied as uncracked $As_4$, and the Sb species as $Sb_2$ molecules. For the fabrication of SAE nanopatterns, commercial single-side polished 2-inch p-type Si (111) wafers were used as substrates, which were covered by a thermal $SiO_2$ mask layer (≈20 nm-thick). Patterns of periodic hole opening arrays were written on the $SiO_2$ mask layer employing electron beam lithography (EBL), reactive ion etching (RIE), and wet chemical etching (buffered hydrofluoric (HF) acid).

Using the prepatterned hole arrays, fully non-catalytic NWs were grown by the growth procedures established in our earlier work.[1-4] First, GaAs(Sb) NWs were grown for 60 min at a substrate temperature of 630°C using Ga flux of 0.35 Å/s, As-BEP (beam equivalent pressure) of $5.5\times10^{-5}$ mbar, and Sb-BEP of $3\times10^{-7}$ mbar. These catalyst-free GaAs(Sb) NWs, which contain only a small Sb molar fraction (≈3–4%), exhibit improved morphological, microstructural, and optical properties due to the so-called Sb-surfactant effect,[3, 4] and therefore serve as stems.

For axially embedding InGaAs-based segments on top of the GaAs(Sb) NW stems, an In-flux of 0.30 Å/s was applied while keeping the Ga- and As-fluxes identical to those used during the NW stem growth, with the substrate temperature ramped down to 590 °C.[5] In this work, the Sb-BEP was intentionally maintained at the same low level ($3\times10^{-7}$ mbar) as during the stem growth to suppress twin formation within the active segment, thereby enhancing the likelihood of well-defined axial growth. Finally, $Al_{0.3}Ga_{0.7}As$ (≈5 nm) / GaAs (≈3 nm) passivation layers were coaxially grown at a substrate temperature of 500 °C.[6]

*Structural Analysis*

The NW growth yield and morphology were evaluated by scanning electron microscopy (SEM) using an NVision 40 FIB-SEM system (Carl Zeiss). Images were acquired at a 45° bird-eye view and tilt-corrected by the SEM software to reflect the actual lengths. High-angle annular dark-field scanning transmission electron microscopy (HAADF-STEM) was performed along the active InGaAs region to characterize the microstructural properties of the NW heterostructures. For these measurements, NWs were mechanically transferred from the SAE array fields onto carbon-coated copper grids and investigated using a FEI Titan Themis TEM operating at 300 kV. Furthermore, energy-dispersive X-ray spectroscopy (EDXS) and corresponding elemental mapping were carried out to analyze the compositional profiles.

*Optical characterizations*

To identify the spatial origin and emission characteristics of the InGaAs(Sb) axial NW quantum emitters, cathodoluminescence (CL), micro-photoluminescence (μPL), and second-order photon-correlation $g^{(2)}(\tau)$ measurements were performed, as schematically described in **Figure S1**. For CL measurements (a), the NW samples are mounted on a the cold finger of a helium flow cryostat mounted to a high-precision interferometric stage and operated at a temperature of 20 K. Excitation is achieved using a 5 kV electron beam delivered through a 30 μm beam aperture. The emitted photons are collected by a high-numerical aperture parabolic mirror and directed onto the entrance slit (50 μm) of a monochromator. A 300 lines/mm diffraction grating, in combination with a silicon CCD detector, is used to spectrally resolve and record the CL signal on a pixel-by-pixel basis, with an integration time of 50 ms per pixel. Secondary-electron (SE) images were acquired concurrently with the CL measurements to correlate the luminescence with the NW morphology.

The setup for μPL and time-resolved PL (TRPL) measurements (b) consists of a closed-cycle cryostat equipped with a three-axis piezoelectric stage, enabling precise spatial positioning of the sample with nm resolution. Optical excitation is provided by a 80 MHz pulsed laser source via a high-NA (0.62) lens, while the emitted PL is analyzed using a monochromator fitted with a 600 lines/mm grating and detected using a silicon CCD. For TRPL measurements, no deconvolution with the instrumental response function was required, as the decay times are substantially longer than the setup time resolution (≈20 to 30 ps). Photon correlation measurements were performed using a Hanbury Brown and Twiss (HBT) interferometer. The setup incorporates a single-mode fiber beam splitter to divide the emission into two detection paths, each coupled to a superconducting nanowire single-photon detector (SNSPD). The arrival times of photons are recorded using a time-correlated single-photon counting (TCSPC) module, allowing determination of the second-order autocorrelation function, $g^{(2)}(\tau)$. We extract $g^{(2)}(0)$ by comparing the counts that contribute to the center peak at zero-time delay with the average counts for uncorrelated peaks.

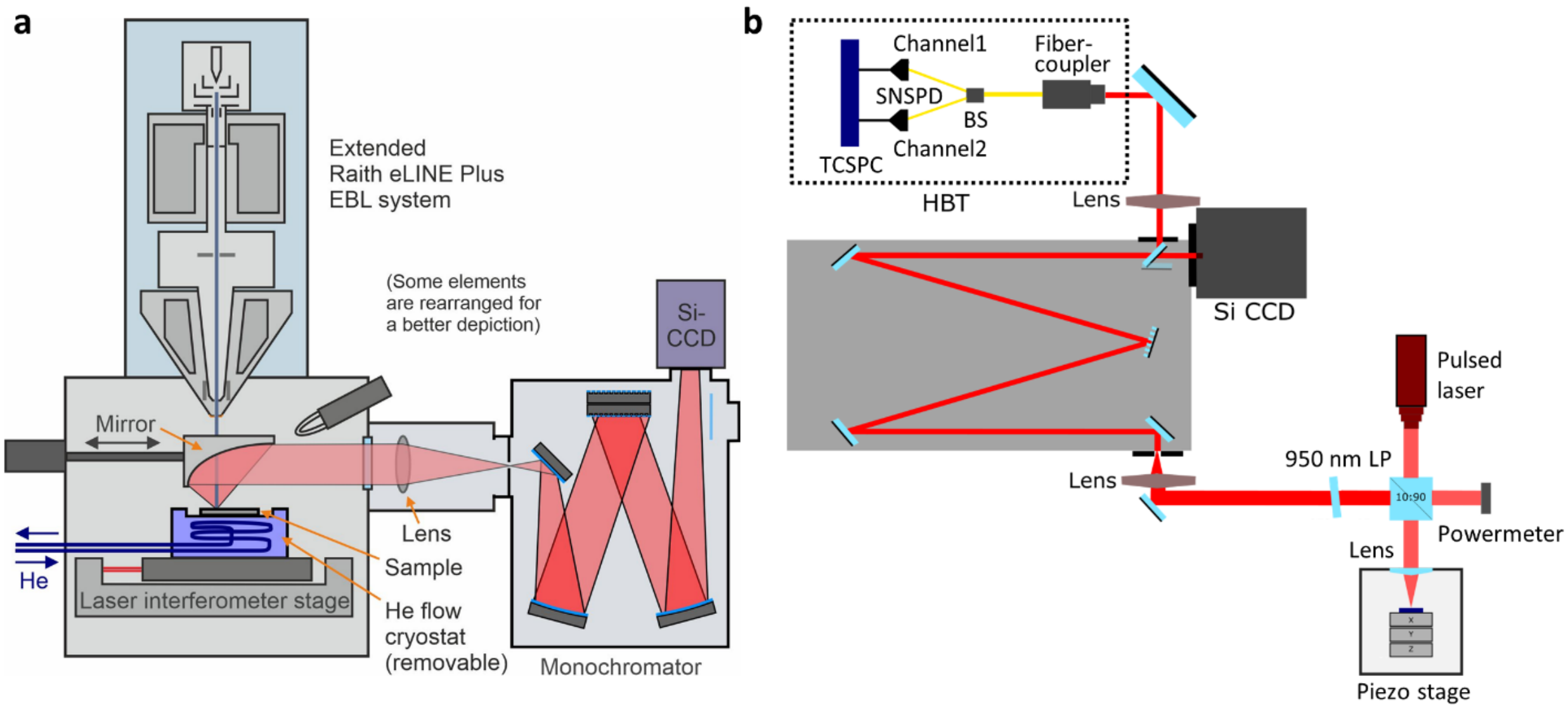

**Figure S1:** Schematic illustrations of the (a) CL and (b) μPL setups used in this work.

## S2. Axial deposition of Sb-free InGaAs

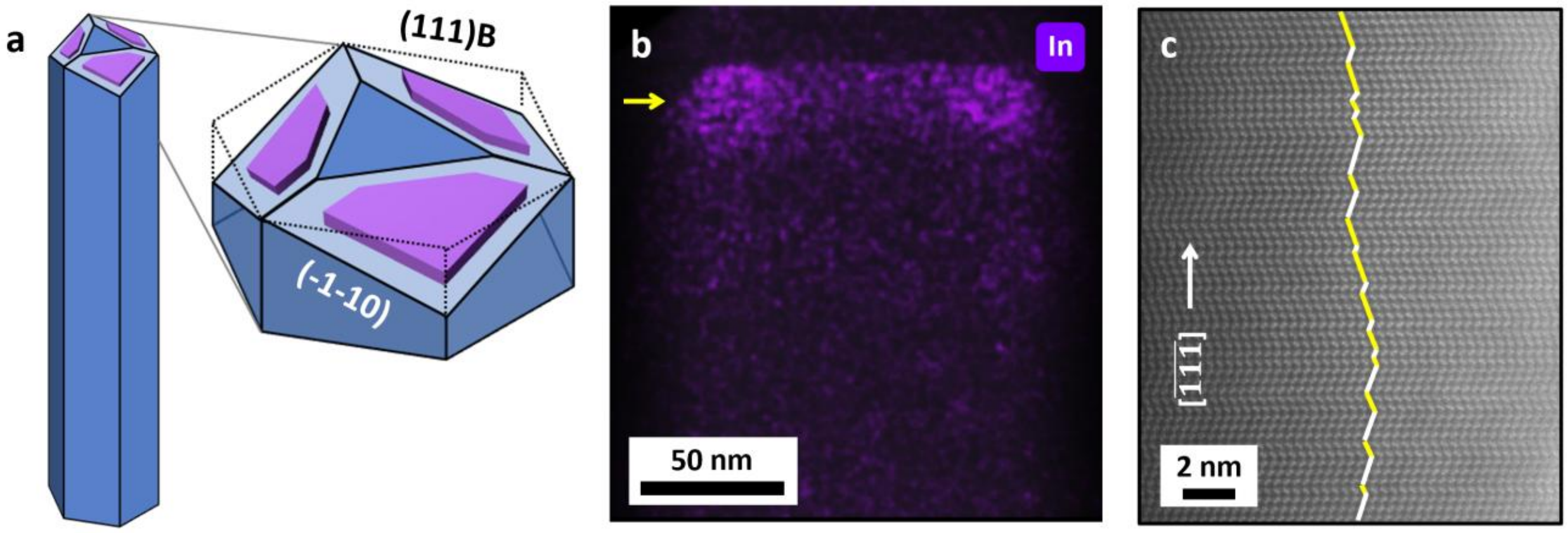

**Figure S2:** (a) Schematic illustration of axial InGaAs deposition on top of a GaAs(Sb) NW stem, showing that Sb-free InGaAs preferentially grows on the undesired {-1-10} inclined facets rather than on the intended (111)B top facet. (b) Elemental distribution map of In (purple), recorded by EDXS along the InGaAs region, indicating dominant deposition on the inclined facets. (c) High-resolution HAADF-STEM micrograph magnifying the InGaAs region (indicated in (b) by the yellow arrow), where alternating rotational twin domains are outlined in yellow and white (twin density ≈1.5 $nm^{-1}$). These observations indicate that the frequent formation of twin defects leads to increased growth on the inclined facets. Additionally, the absence of Sb surfactant reduces the stability of the (111)B growth front, further suppressing axial disk formation on the top facet.

## S3. Additional features of axial InGaAs(Sb) - stochastic nature

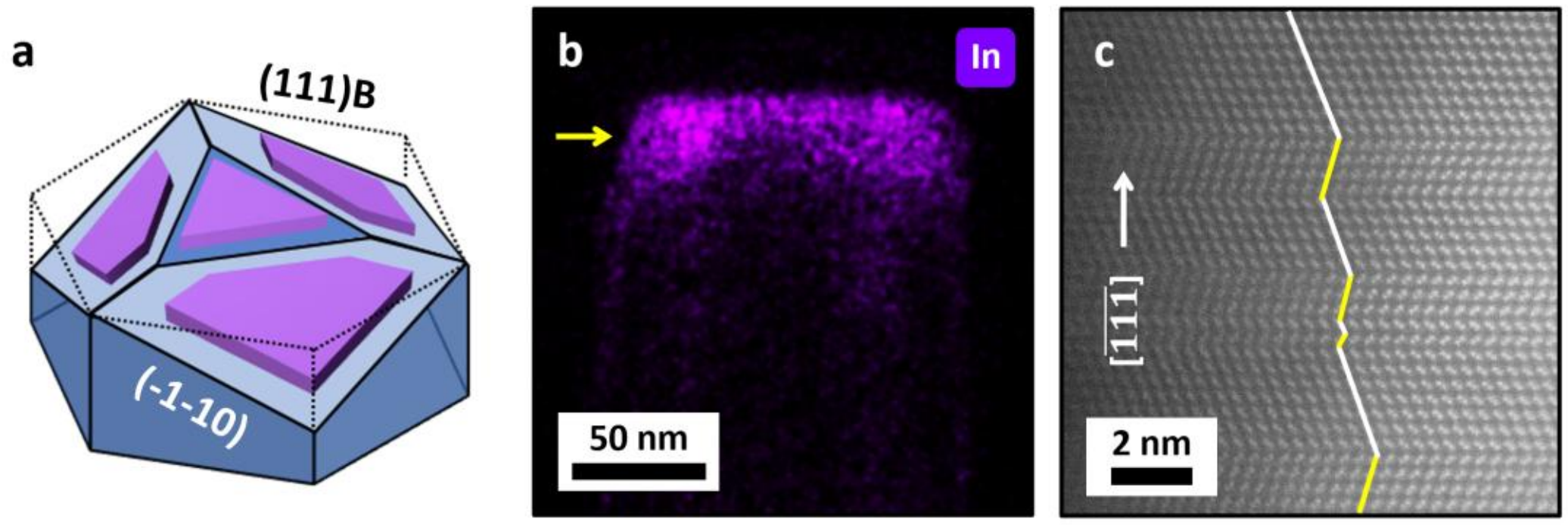

**Figure S3:** (a) Schematic illustration of axial InGaAs(Sb) deposition on top of a GaAs(Sb) NW stem, showing that, due to its stochastic nature, Sb-containing InGaAs(Sb) grows not only on the intended (111)B top facet but also more dominantly on the undesired {-1-10} inclined facets. (b) Elemental distribution map of In (purple), recorded by EDXS across the InGaAs(Sb) region, indicating deposition on both the top and inclined facets. (c) High-resolution HAADF-STEM micrograph magnifying the InGaAs(Sb) region (indicated in (b) by the yellow arrow), where alternating rotational twin domains are outlined in yellow and white, with a ≈three-fold reduced twin density (≈0.5 $nm^{-1}$) compared to the Sb-free InGaAs sample (≈1.5 $nm^{-1}$). These observations indicate that the presence of twins is directly linked to inclined-facet growth. Although Sb surfactant reduces the twin density and stabilizes growth on the (111)B top facet, the stochastic nature of twin formation still leads to a subset of NWs within the array exhibiting unintended growth on the inclined facets. Note that twin formation may not redirect material from the (111)B top facet but rather activates additional incorporation channels on the inclined facets, leading to a larger overall InGaAs(Sb) volume compared to twin-free segments where growth remains confined to the top facet.

## S4. Estimation of quantum-confined InGaAs(Sb) bandgap energy

First, assuming a bulk InGaAsSb alloy with [In] = $(20 \pm 2)$% and [Sb] = 4%, and using a standard Vegard's law interpolation combined with bowing parameters for the InGaAs[7] and InAsSb sub-systems[8, 9], the estimated bandgap energy at low temperature (0 K) lies in the approximate range

$$E_g^{bulk} \approx 1.18 – 1.23 \text{ eV.}$$

In addition, the bandgap modification induced by quantum confinement can be roughly evaluated using the effective-mass approximation ($m_e^* \approx 0.055\text{m}_0$ and $m_{hh}^* \approx 0.49m_0$ for $In_{0.2}Ga_{0.8}As$ in the ZB structure[10]) and a rectangular quantum box model (infinite potential well). For a simplified rectangular quantum structure with lateral dimensions $L_x = L_y \approx 80$ nm and an active-layer thickness of $L_z \approx 7–10$ nm (effective confinement width accounting for compositional grading), the electron and heavy-hole confinement energies can be estimated as follows[11].

For the axial direction (here, $L_z$ = 7 nm),

$$E_{e,z} = \frac{\hbar^2\pi^2}{2m_e^* L_z^2} \approx 0.139 \text{ eV}, \qquad E_{hh,z} = \frac{\hbar^2\pi^2}{2m_{hh}^* L_z^2} \approx 0.0156 \text{ eV.}$$

For the lateral direction ($L_x = L_y \approx 80$ nm),

$$E_{e,xy} = \frac{\hbar^2\pi^2}{2m_e^*}\left(\frac{1}{L_x^2} + \frac{1}{L_y^2}\right) \approx 0.00213 \text{ eV}, \qquad E_{hh,xy} = \frac{\hbar^2\pi^2}{2m_{hh}^*}\left(\frac{1}{L_x^2} + \frac{1}{L_y^2}\right) \approx 0.00024 \text{ eV.}$$

The total increase in the interband transition energy due to quantum confinement is therefore

$$\Delta E \approx \left(E_{e,z} + E_{e,xy}\right) + \left(E_{hh,z} + E_{hh,xy}\right) \approx 0.157 \text{ eV.}$$

The resulting quantum-confined bandgap energy ($E_g^{Conf}$) is estimated to be

$$E_g^{Conf} \approx E_g^{bulk} + \Delta E \approx 1.33–1.39 \text{ eV}$$

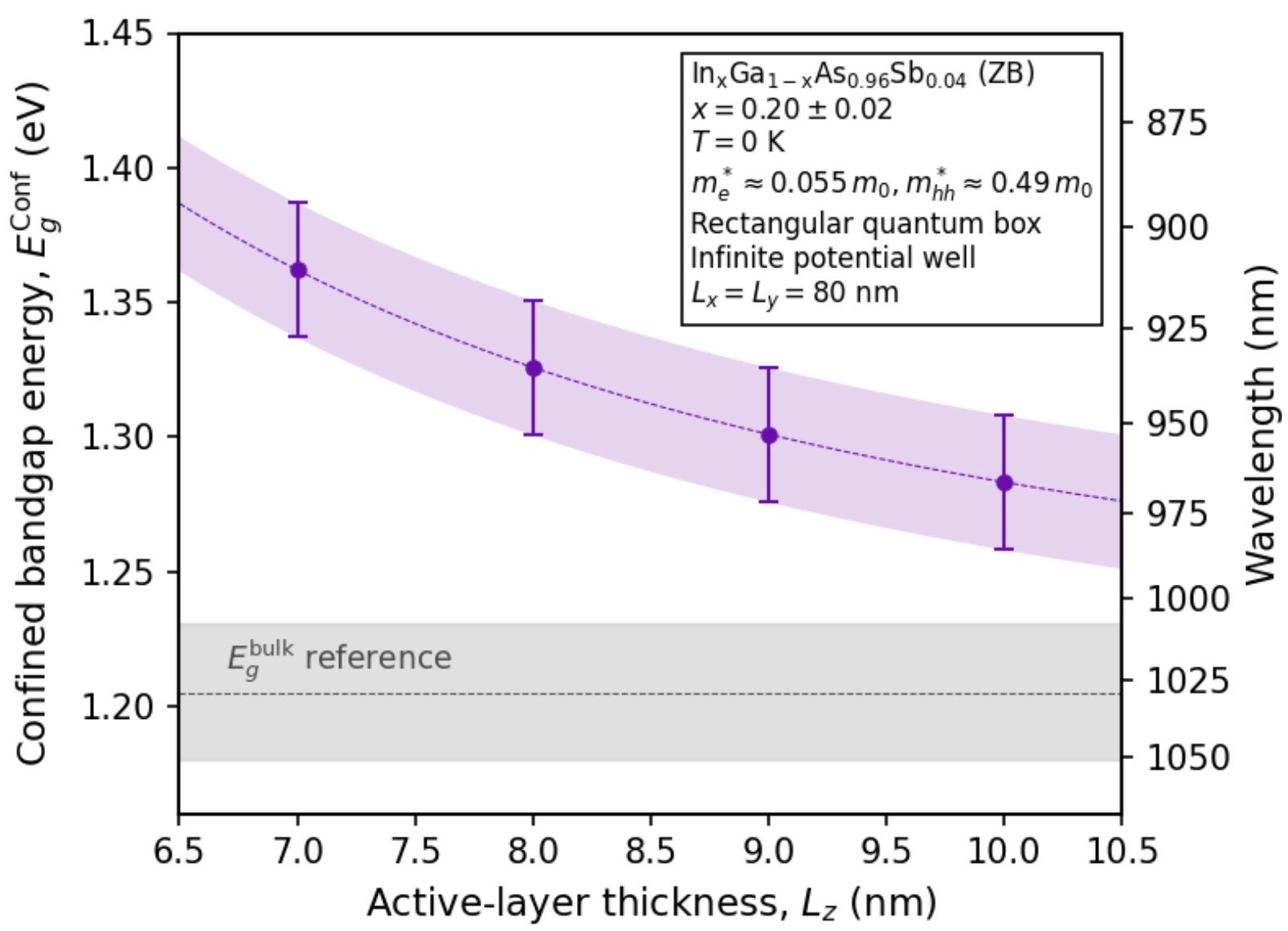

**Figure S4:** Estimated confined ground state energy ($E_g^{Conf}$) as a function of active layer thickness (purple), together with the corresponding bulk bandgap energy ($E_g^{bulk}$, gray) for reference.

Thus, for quantum structures with a thickness of $L_z \approx$ 7–10 nm and lateral dimensions of $L_x = L_y \approx 80$ nm, the confined ground state energy is estimated to lie in the range

$$E_g^{Conf} \approx 1.25\text{–}1.39 \text{ eV},$$

corresponding to wavelengths of $\lambda \approx 890\text{–}990$ nm, as summarized in **Figure S4**.

Although additional bandgap shifts arising from strain within the NW, variations in twin defect density, and local compositional (both In and Sb) or structural fluctuations should ideally be taken into account, this order-of-magnitude estimate nevertheless yields transition energies comparable to the typical range of experimentally observed PL peak energies. It should also be noted that this model assumes an infinite potential well and fixed effective masses, therefore, the calculated value likely represents an approximate upper limit on the confinement-induced blue shift relative to the bulk bandgap.

## S5. Additional CL data

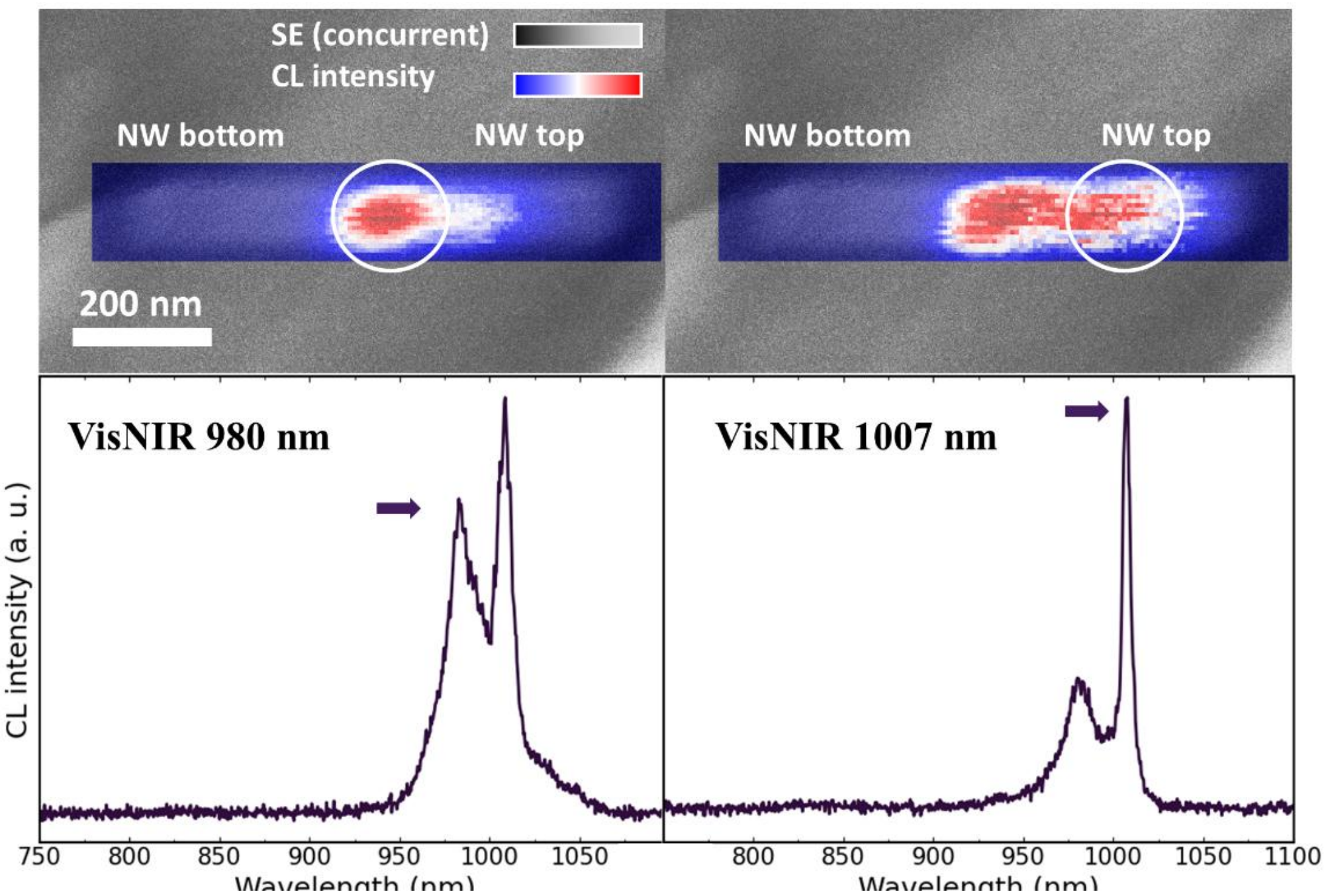

**Figure S5:** CL intensity maps recorded using spectral windows centered at 980 nm (left) and 1007 nm (right), along with the corresponding CL spectra extracted from the same NW. The emission detected around 980 nm is mainly localized near the central region of the NW, whereas the 1007 nm emission, corresponding to the spectral range where pronounced antibunching with $g^{(2)}(0)$ of 0.38±0.02 was observed in μPL, is concentrated closer to the NW tip region. These data provide supportive evidence that the single-photon emission is associated with the InGaAs(Sb) insertion near the NW tip, although the two spectral contributions cannot be fully spatially isolated in the CL maps due to the spectral overlap of these emitters.

## S6. Additional second-order correlation measurements

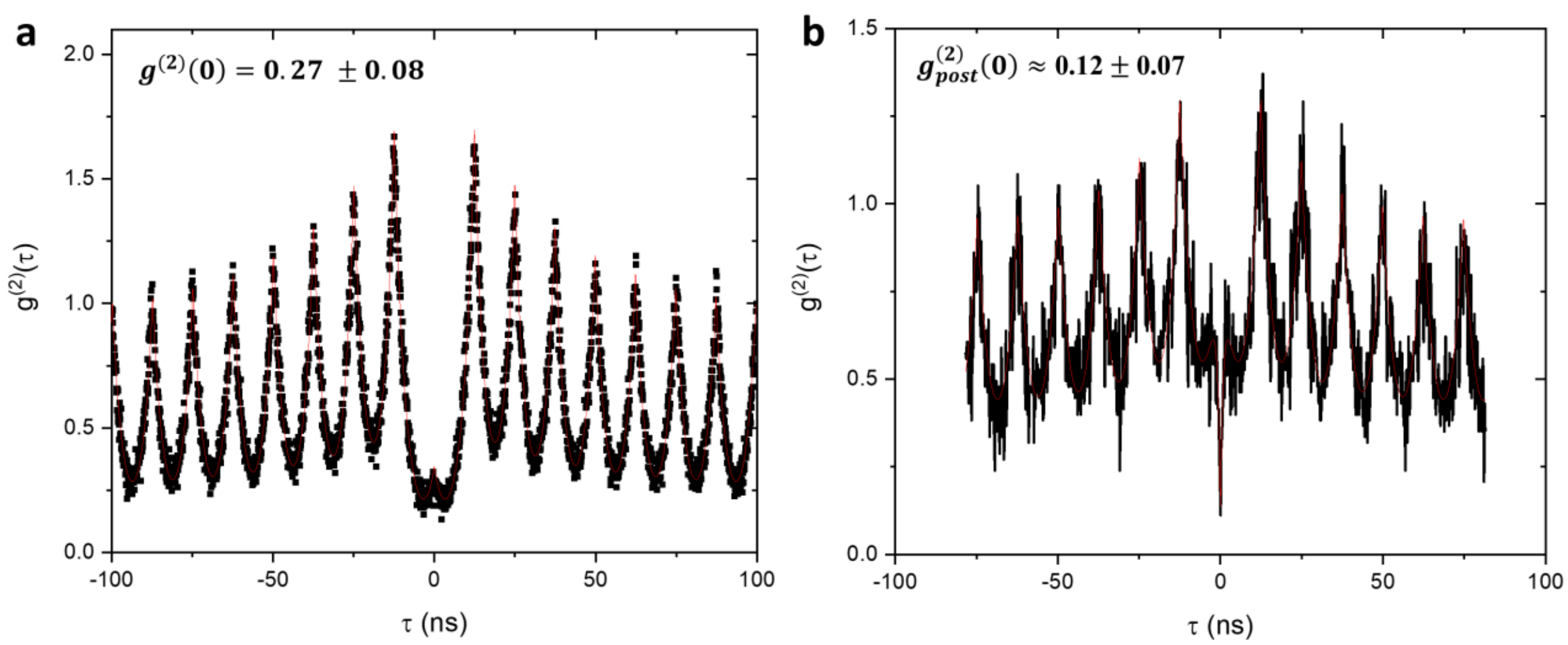

**Figure S6:** (a) Second-order photon-correlation function $g^{(2)}(\tau)$ measured under pulsed excitation, yielding a raw $g^{(2)}(0)$ value of 0.27±0.08. The suppressed zero-delay peak confirms antibunching, while the weak bunching envelope ($g^{(2)}(\tau) > 1$) extending over multiple pulse periods suggests the presence of additional slow emitter state dynamics, such as charge trapping, recapture, or blinking-related effects. (b) $g^{(2)}(\tau)$ data from another emitter, yielding $g^{(2)}(0)$ = 0.12±0.07 after temporal post-selection. In this case, the long radiative lifetime leads to substantial overlap between neighboring peaks, causing a pronounced background around zero delay. Therefore, the post-selected value corresponds to the minimum at zero delay, such that only photons emitted immediately after the excitation pulse are counted, while delayed emission associated with recaptured carriers is excluded.

## REFERENCES FOR SUPPORTING INFORMATION